# Laser-like emission from a periodic all-silicon nanostructure

Sylvain G. Cloutier[1] and Jimmy Xu[1,2]
[1] Division of Engineering, Brown University, Providence, RI 02912, USA
[2] Department of Physics, Brown University, Providence, RI 02912, USA
(Submitted: Nov. 8th, 2004)

We report on experimental observation of laser-like emission from a periodic all-silicon nanostructure formed on an electronic grade silicon-on-insulator wafer, using a highly-uniform hexagonal self-organized array of nanopores as an etch mask. Classical lasing characteristics such as threshold behavior and stimulated emission, optical gain, emission linewidth narrowing and far-field modal pattern were observed at temperatures below 70 K under optical pumping.

Although highly desirable, a silicon laser has not been thought feasible. This general belief is rooted in the fundamental fact that crystalline silicon is an indirect bandgap semiconductor in which radiative recombination of electrons and holes is highly unlikely due to the large mismatch in the k-space between the available electron and hole states and competition with a myriad of faster non-radiative processes. Despite the fundamental limitations on light emission in silicon, attempts have been made, especially over the last two decades, to make silicon a more efficient light emitter [1-10]. Many of these efforts aimed at enhancing silicon light emission from defect-states or surface recombination via ion implantation [4,5,11], chemical impregnation [6], improved external coupling of the emitted photons [3], or quantum-confinement effects [2,5,7-10]. Whereas enhanced and amplified light emission has been achieved to varying degrees in the prior efforts of using porous silicon, silicon nano-particles, or other techniques, no lasing action was reported [1]. Theoretically, optical gain is possible in indirect semiconductor [12,13]. Optical gain and stimulated emission through excitonic molecules [14] or isoelectronic trapping [15] have been experimentally observed in other indirect-bandgap semiconductors and more recently claimed in silicon nanocrystals [2,10] and ion-implanted silicon [5]. In this Letter, we report on laser-like emission characteristics obtained in a periodic all-silicon nanostructure.

Shown in Fig.1(a), this all-silicon mesoscopic structure was fabricated by etching a highly ordered array of nanosize holes (antidots) into a crystalline silicon layer on an electronic grade silicon-on-insulator (SOI) wafer.

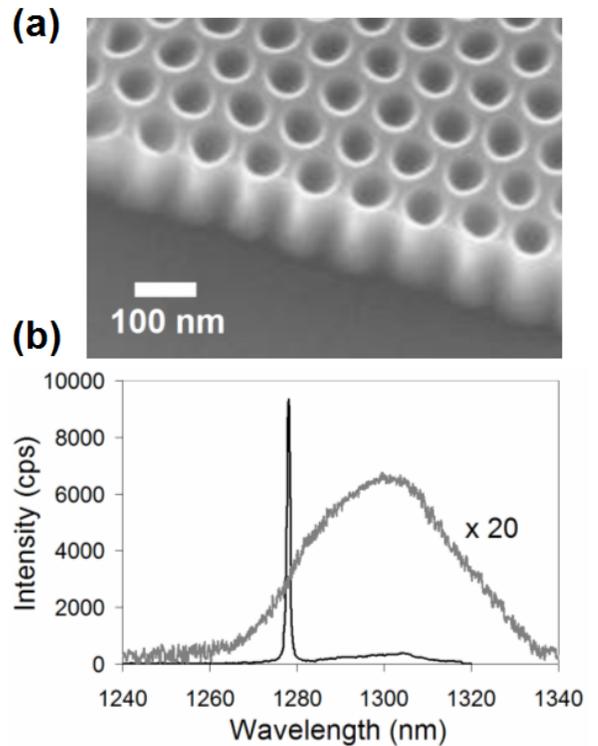

FIG. 1. (a) 45-degree view of a typical nano-patterned silicon-on-insulator sample under scanning-electron microscope. (b) Edge-emission spectra at 10 K for 60 and 600 mW/mm$^2$ pump-power densities.

The original SOI wafer has a 205nm thick undoped silicon <100> layer atop 3μm of oxide on a thick silicon substrate. We first thinned the top silicon layer of such an SOI wafer to 65 ± 5 nm via thermal-oxidation followed by a highly selective wet-etching of the oxidized silicon in a 1:6 HF:$H_2O$ solution. The thinning was performed in an attempt to bring all the prime feature sizes into the same range. One consequence of thinning the active layer so much is that the absorbed power would only be a very small fraction of the apparent pump power as can be estimated from the product of the film thickness and the absorption coefficient. One benefit is in that the emitted light would extend more into less absorptive media. Un-patterned pieces from the same thinned SOI sample were set aside for control reference and comparison.

This high degree of uniformity in the nanopatterned structure is enabled by the use of the nanopore array anodic aluminum oxide (AAO) membrane as an etch mask. The AAO mask was made using a two-step anodization process in which uniform nanopores form and self-organize into a highly-ordered hexagonal array. The pores diameter and spacing can be controllably changed under carefully controlled anodization conditions [16,17]. The AAO membrane used here was formed under the conditions previously described [16,17] up to a 750nm thickness, and subsequently placed atop the SOI wafer while immersed in water.

Finally, the nano-pore array pattern was transferred into the silicon using chlorine-based reactive-ion etching (RIE). The AAO acts as an effective mask in the process and permitted the formation of clean, deep and straight nano holes in the top silicon layer. After RIE, the AAO mask was removed. This technique therefore allowed controlled nano-patterning over a relatively large area. With this pattern-transfer approach, we are also able to keep the process as simple and the fabricated sample as clean as possible.

Samples used for experiments had resultant pores with 50nm in diameter and 110nm in spacing. 1.00 ± 0.03 mm-wide slivers of such patterned SOI were cleaved for edge-emission measurements and compared with non-patterned thinned-SOI. To keep internal material loss at minimum, we chose undoped crystalline SOI for the experiments which therefore required optical pumping.

The samples were loaded in a cryostat for cooling and a continuous-wave 514.5nm Argon-ion laser beam was focused onto the sample's top-surface, the edge-emission being collected and focused on a TRIAX spectrometer entrance slit for spectral measurements. As shown in Fig.1(b), the edge-emission at 10 K from the nano-patterned sample exhibited a broad emission band from 1260 to 1340 nm at low pump-power (60 mW/$mm^2$). Upon raising the pump power level to 600mW/$mm^2$, a sharp emission peak emerged at 1278 nm, with a full-width half-maximum (FWHM) under 0.9 nm. The lasing peak persisted with increasing temperature up to 70K. No such emission-bands (broad or sharp) were observed from the reference non-patterned samples.

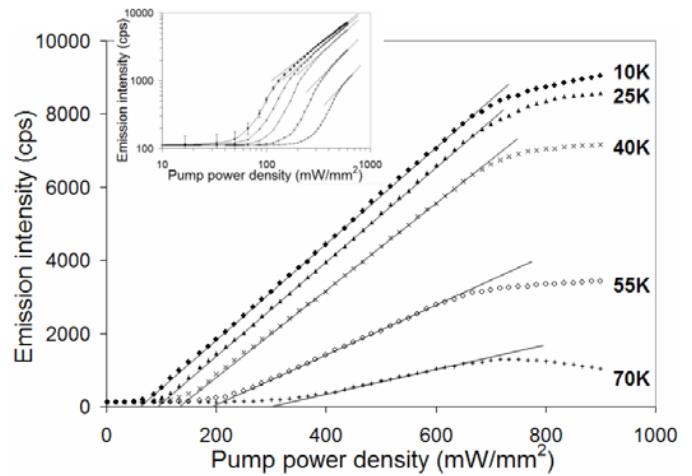

**FIG. 2.** Evolution of the edge-emission 1278nm-peak intensity as a function of the incident pump power for the device operating at 10, 25, 40, 55 and 70 K. The inset shows the data on logarithmic scales to highlight the threshold transition. The excited area is 0.2 $mm^2$.

The peak power as a function of the incident pump power is shown in Fig.2 for different operation temperatures (10, 25, 40, 55 and 70 K). It shows a clear threshold behavior, evolving from sub-threshold emission to a linear output region then reaches apparent saturation. Note that the power-evolution change with temperature follows what is expected from classical semiconductor lasers with a thin active layer [18]. The inset displays the sub-threshold transition behavior on logarithmic scales. This transition from sub-threshold to over-threshold behaves also as expected from an ideal laser [19] and is consistent with prior observations in conventional semiconductor lasers [20].

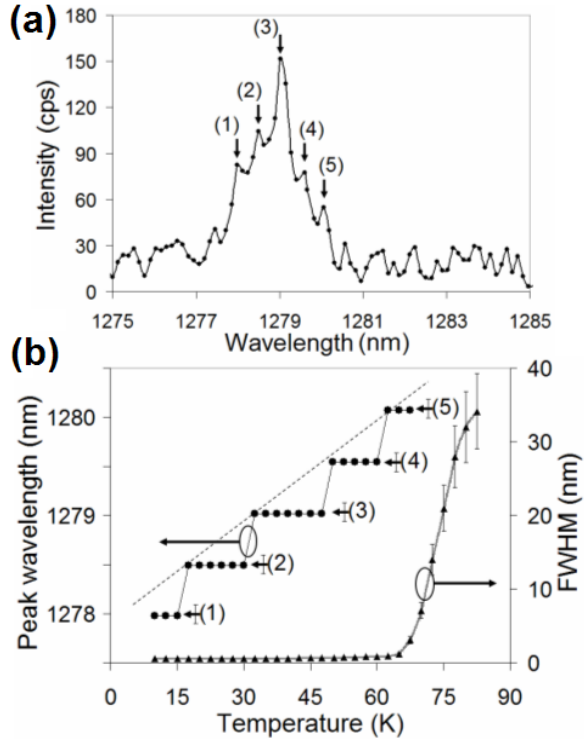

**FIG. 3.** *(a) High-resolution edge-emission spectra at 750 mW/mm2. (b) Evolution of the edge-emission peak wavelength and full-width half-maximum as a function of the operation temperature.*

Moreover, Fig.3(a) shows a high-resolution spectral measurement which reveals multiple spectral peaks typical of resonant-cavity modes. In Fig.3(b), the existence and spectral positions of those peaks are confirmed from the stair-case evolution of the peak wavelength with increasing temperature. The measured frequency spacing between these modes is 96 ± 1 GHz, which corresponds to an effective optical cavity length of 1.563 ± 0.015 mm (defined as the product of $N_{eff}$ x L, with $N_{eff}$ being the guided modal effective refraction index and L the physical cavity length). The modal calculation for the waveguiding structure is shown in Fig.4(a). A 3.06 average material refraction index, calculated for 18.7% air-pores and 81.3% bulk silicon (n = 3.54 at 1278 nm), was obtained for the 65 nm patterned silicon layer. From there, one can find the guided modal effective index to be 1.564 at a 1278 nm wavelength for such 65 nm-thick nanopatterned structure with air-cladding on one side and amorphous $SiO_2$ on the other. From the modal effective index, one finds that the measured modal-spacing of Fig.3(a) and Fig.3(b) corresponds to a physical cavity length of L = 1.00 ± 0.02 mm, matching perfectly the measured cleaved-sample width (1.00 ± 0.03 mm) confirming longitudinal resonant cavity feedback modes.

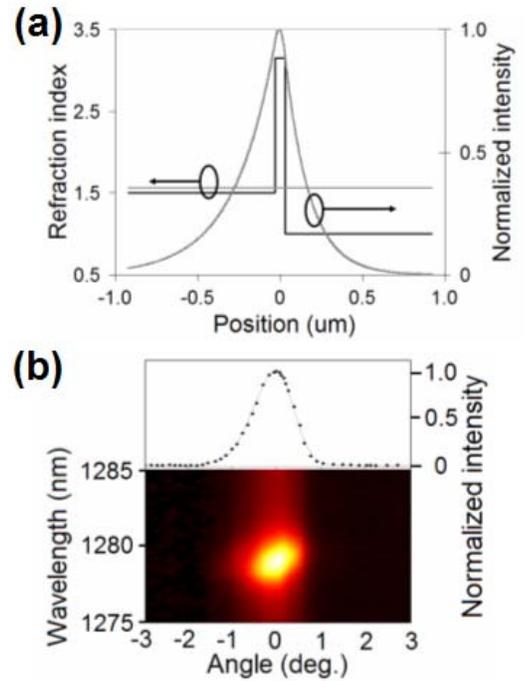

**FIG. 4.** *(a) Waveguide-structure refraction index profile and calculated modal intensity distribution. The calculated modal effective index is 1.564 (b) Lateral angular distribution of the emitted light at 10 K and 450 mW/mm$^2$. The top panel shows the angular distribution of the 1278 nm emission peak while the bottom one shows the spectral angular-distribution.*

From the emission-peak FWHM evolution with temperature shown in Fig.3(b), one sees that,

at a given pump power, the device ceases lasing at around 72.5 K, consistent with the measured threshold behaviour in Fig.2. Beyond this point, the emission spectrum broadens rather abruptly.

In addition to the spectral narrowing, we also made the effort to observe the far-field modal distribution – another classical indicator of lasing. This is made difficult by the low total emission power corresponding to the very small active (gain) volume. Nevertheless, we were able to measured the lateral far-field intensity distribution clearly, which is shown in Fig.4(b) and displays a FWHM of 1.1 ± 0.1 degree, as expected for spatially-coherent emission from a wide and thin stripe of active material [18,19].

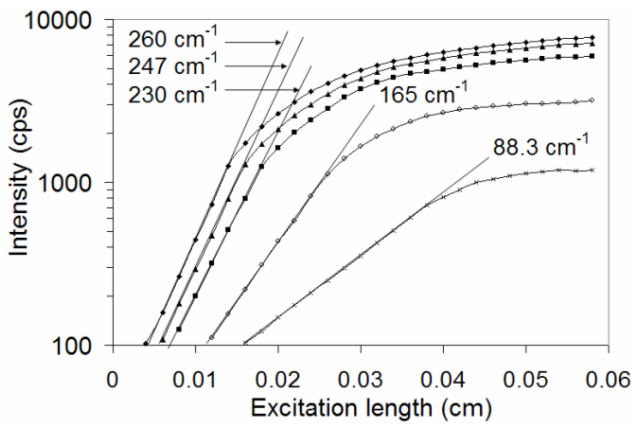

**FIG. 5.** *Evolution of the edge-emission 1278nm-peak intensity as a function of the excited length at 10, 25, 40, 55 and 70 K. The excitation was 650 mW/mm$^2$.*

Finally, using the well-established variable stripe length (VSL) method [21], we measured the evolution of the peak-emission intensity as a function of the sample's excitation length. The results are displayed on a semi-logarithmic scale in Fig.5. The linear-evolution region on this semi-log plot represents an expected exponential variation of intensity, a clear evidence of optical gain [22,23]. The gain coefficient can be obtained from the slope of the linear region (using an exponential fit), and it reads 260, 247, 230, 165 and 88 ± 20 cm$^{-1}$ at 10, 25, 40, 55 and 70 K respectively.

These experimental observations are strongly indicative of lasing action. The presence of a significant optical gain in the periodically nanostructured silicon meets one of the first essential conditions for amplified stimulated emission and lasing action [14,18,19,22,23]. The threshold behavior of Fig.2 is a first indicator of stimulated emission and population inversion. The spectral line narrowing with increasing pump power and narrow far-field angular distribution, the presence of cavity mode peaks in the high-resolution emission spectra are further evidences consistent with lasing action and coherence [18,19,24].

Whereas we are pleased to have obtained these results and to report them as facts that all point to the occurrence of lasing action in this all-silicon nanostructure, we are not yet in position to offer any convincing explanation of the mechanism(s) by which the laser or laser-like action has occurred. At this very early stage, explanations of the underlying physics, if any, would necessarily be speculative. It is also entirely possible that the system may be operating in a new regime previously inaccessible but now opened by some unique attributes in this periodically nanostructured all-silicon system. Nevertheless, it seems to be constructive, to future explorations that are likely to be long-term and broad-ranging in nature and we hope this report will help launch, to speculate on a few possible contributing factors. The nanoscale patterning may have induced localized symmetry breaking and strain in the side wall of each cylindrical hole, and in so doing, have altered the local lattice and band structure to permit enhanced light emission. The occurrence of the peak emission at a wavelength below the silicon band edge is suggestive of that possibility. With high uniformity, the scattering and material loss and the inhomogeneous broadening can be kept at minimum, thereby permitting a significant net optical gain in this material, as evidenced in Fig.5. Theoretically it has been shown that it is possible to have optical gain in indirect bandgap semiconductors (e.g. silicon) exceeding the material loss [12, 13]. With a finite net material gain and a sufficiently long cavity, lasing actions are indeed permitted.

At the cryogenic temperatures, exciton interactions could play a dominant role and lead to stimulated emission [24-26]. A high concentration of excitons is expected at the pumping levels deployed in the experiments reported here. The nano-structured side-wall regions may help localize and condense the excitons. Both the thin crystalline silicon on the oxide and the presence of a thick electronic grade oxide layer help to preserve the excitons and extend their lifetime. Early on, we also considered electron-hole droplets [25] a possibility when the lasing action was observed only at sub-30K temperatures, which were subsequently ruled out when we achieved lasing at much higher temperatures as presented here. Anderson photon localization [27] is another effect that could not yet be excluded but seems to be less likely.

In summary, we present experimental findings of laser-like emission from a highly uniform periodic all-silicon nanostructure. Classic features of laser emission such as optical gain, stimulated emission, spectral linewidth narrowing and narrow far-field intensity pattern indicative of coherence, and threshold behavior were all clearly present in the edge-emission measurements. It is also informative to add that the emitted light was unpolarized below threshold and more than 80% polarized, parralel to the pores axis, (vertical to the plane) above threshold and that subsequent measurements on additional samples made independently but using the same recipe by different members of the lab yielded similar results. We note that the pump power level may appear to be high when compared to the emitted power level, however, the actual absorbed power is only a small fraction because of the extremely small optical density of the 60nm-thin and porous active layer. While the observed lasing action is so far only at cryogenic temperatures, further developments and optimization of the structure guided by a good understanding of the underlying physics can be expected, and may likely enable improved performances, operation at higher temperatures and under electrical pumping conditions.

We acknowledge timely and enabling support from ONR and DARPA, and the help and suggestions from Pavel Kossyrev, Hope Chik, and Martin Fay. S.G.C. is also thankful to NSERC for its support.